\documentclass[a4paper]{article}
\usepackage{parskip}
\usepackage{pslatex}
\usepackage{graphicx}
\usepackage{amsmath}
\usepackage{amsfonts}
\usepackage{amssymb}
\usepackage[utf8]{inputenc}
\usepackage{mathtools}
\usepackage{algorithm}
\usepackage{enumitem}
\usepackage{upgreek}
\usepackage{tikz}
\usepackage{bibentry}

\usepackage{authblk}
\providecommand{\keywords}[1]
{
  \small
  \textbf{\textit{Keywords---}} #1
}

\newcommand{\Reynolds}{\mathrm{Re}}
\newcommand{\Weber}{We}
\newcommand{\Kapitza}{Ka}
\newcommand{\rd}{\Pi_{\rho}}
\newcommand{\rv}{\Pi_{\mu}}
\newcommand{\uN}{u_N}
\newcommand{\hN}{h_N}

\newcommand{\set}[1]{{#1}}                  
\newcommand{\dscl}[2][s]{\mathbf{{#2}_{#1}}}      
\newcommand{\dvec}[1]{\mathbf{#1}}
\newcommand{\Id}[1]{\mathbb{I}_{#1}}        
\newcommand{\Ic}[1]{\mathsf{I_{#1}}}        
\newcommand{\Zv}[1]{\mathbf{0_{\set{#1}}}}  
\newcommand{\sfield}[2][]{\dscl[#1]{#2}}     
\newcommand{\vfield}[1]{\dvec{#1}}    
\newcommand{\DOT}[3][]{\left ( #2, #3 \right )_\mathsf{#1}}
\newcommand{\strain}{S}                     
\newcommand{\stress}{\sigma}                
\newcommand{\st}{\gamma}                    
\newcommand{\vel}{\vec{u}}                  

\newcommand{\ki}[1][]{\kappa_\set{#1}}      
\newcommand{\kc}{\sfield[c]{k}}             
\renewcommand{\ni}{\hat{\eta}_{i}}          
\newcommand{\ls}{\theta}                    
\newcommand{\lsv}[1]{\sfield[#1]{\uptheta}} 
\newcommand{\LS}[1][C]{\sfield[C]{\Theta}}  
\newcommand{\grad}{\nabla}                  
\renewcommand{\div}{\nabla \cdot}           
\newcommand{\RHO}[1][]{\mathsf{P_{#1}}}
\newcommand{\DIV}[1][]{\mathsf{D}_{\mathsf{#1}}}     
\newcommand{\GRAD}[1][]{\mathsf{G}_{\mathsf{#1}}}    
\newcommand{\LAP}[1][]{\mathsf{L}_{\mathsf{#1}}}     
\newcommand{\CONV}[1][]{\mathsf{C}(\uf)_{\mathsf{#1}}}
\newcommand{\ADV}[2][]{\mathsf{C}(#2)_{\mathsf{#1}}}
\newcommand{\OFFADV}[2][]{\mathsf{C}^{off}(#2)_{\mathsf{#1}}}
\newcommand{\U}[1][]{\mathsf{U}}            
\newcommand{\E}{{\Upsilon}}                 
\newcommand{\HR}{\Psi}                      
\newcommand{\sCF}{\Pi}                      

\newcommand{\pc}{\sfield[c]{p}}             
\newcommand{\uf}{\sfield[f]{u}}             
\newcommand{\K}[1][]{\mathsf{K_{#1}}}       

\title{Energy preserving multiphase flows: Application to falling films.}

\author{N. Valle, F. X. Trias and J. Castro}

\affil{
  Heat and Mass Transfer Technological Centre (CTTC), Universitat Polit\`ecnica 
  de Catalunya - BarcelonaTech (UPC), ESEIAAT, Carrer Colom 11, 
  08222 Terrassa (Barcelona),\\ \{nicolas.valle, francesc.xavier.trias, 
        jesus.castro\}@upc.edu}

\nobibliography*
\begin{document}
\maketitle
\begin{abstract}
    The numerical simulation of multiphase flows presents several challenges, namely 
the transport of different phases within de domain and the inclusion of 
capillary effects.
Here, these are approached by enforcing a discrete physics-compatible solution.
Extending our previous work on the discretization of surface tension 
[\bibentry{Valle2020}] with a consistent mass and momentum transfer
a fully energy-preserving multiphase flow method is presented.
This numerical technique is showcased within the simulation of a falling film 
under several working conditions related to the normal operation of 
$\mathrm{LiBr}$ absorption chillers.

\end{abstract}
\section{INTRODUCTION}
\label{sec_ff:introduction}

Vertical falling films are a canonical flow configuration which is inherently 
unstable even at $\Reynolds = 0$.
When surface tension is present, the dynamics of such a system turn even more 
complex and interesting.
Its application is of interest for many industrial applications in which high 
heat and mass transfer coefficients are expected with low temperature jumps.
These include heat exchangers used in desalination and gas absorbers, like 
$\mathrm{HCl}$ absorbers used in chlorination processes and, most remarkably, 
$\mathrm{H_2O-LiBr}$ absorption chillers.
The ultimate goal is to gain understanding on the instabilities appearing on 
this flow as a keystone to approach the heat and mass transfer processes in 
subsequent steps.
In this regard, while the vapor phase has little effect in the fluid dynamics,
it plays a key role when considering heat and mass transfer, in addition to 
ruling other non-linear phenomena such as the transport of volatile surfactant.

In the computational front, simulations of Nave et al.~\cite{Nave2010}
replicated the experiments of Nosoko et al.~\cite{Park2003} by using a ghost 
fluid - level set method and showed potential to capture flows transitioning 
from $2\mathrm{D}$ to $3\mathrm{D}$. The dominant role of inertia in governing 
the surface waves~\cite{Kalliadasis2012} has been studied by Denner et 
al.~\cite{Denner2016,Denner2018}, who performed remarkable experimental and 
numerical work.
The role of capillary flow separation at increasing heat and mass transfer 
characteristics was assessed by Dietze and Kneer~\cite{Dietze2011a} from both 
experimental and numerical results, bolstering the role of surface tension at 
promoting heat and mass transfer in this flow configuration.
Mass transfer enhancement at the capillary waves region was also confirmed by Bo 
et al.~\cite{Bo2011} in their $2\mathrm{D}$ simulation of a $\mathrm{H_2O-LiBr}$ 
absorber at $\Reynolds = 100$ and by
Garcia-Rivera et al.~\cite{Garcia-Rivera2016}.

Similarly, the work of Albert et al.~\cite{Albert2014} performed numerical 
simulations in order to asses the heat and mass transfer characteristics of 
falling films, which confirmed, again, the role of flow separation for the 
enhancement of such phenomenon.
The study of $3\mathrm{D}$ structures in Dietze et al.~\cite{Dietze2014} also 
observed flow separation.
The assessment of corrugations and its impact in heat and mass transfer 
phenomena was also assessed by Dietze~\cite{Dietze2018} for $\Reynolds=15$.


In this context, the assessment of inertia and capillary terms, which rule the 
flow instabilities that promote heat and mass transfer, is of relevance for the 
development of new $\mathrm{H_2O-LiBr}$ absorbers.
Nonetheless, the adoption of physically inconsistent schemes is customary for 
both convective and capillary terms.
Accentuated by the high density ratios, this leads to inaccurate results and 
even numerical instabilities that compromise the accurate solution of the 
system.
This highlights, once again, the importance of physics-compatible 
discretizations.

The simulation of multiphase flows in a fully physics-compatible way implies the 
conservation of discrete primary quantities (i.e., mass and momentum) and also 
secondary ones (i.e., mechanical energy) according to the physics described in 
the continuum.
Two major obstacles prevent us from this: the inclusion of surface tension and 
the advection of a varying density flow.

In order to tackle these issues, symmetry-preserving ideas \cite{Verstappen2003} 
have been used to set the mathematical grounds of the energy conservation in the 
context of fluid flow simulations. Those provide with a high degree of physical 
reliability, but also with improved stability.

Regarding the inclusion of surface tension, Fuster~\cite{Fuster2013} developed 
on top of a Volume Of Fluid~(VOF) method a discretization focusing on preserving 
the (skew-)symmetry of the operators involved, albeit overlooked surface 
tension, which is know to introduce spurious oscillations that may eventually 
lead to the divergence of the numerical simulation~\cite{Magnini2016}.
In the context of the level set method, our previous work resulted in the 
inclusion of curvature in an energy-preserving fashion~\cite{Valle2020}, while 
the conservation of momentum is still elusive, a well known issue for diffuse 
interface models~\cite{Kim2005}.
Within phase-field methods, the pioneering work of Jacqmin~\cite{Jacqmin1999} 
included surface tension in a consistent way in the context of the Cahn-Hilliard 
equation.
Phase-field methods success at capturing surface tension transfers between 
potential (elastic) and kinetic energy rely on taking the gradient of a surface 
potential, while VOF and level set methods aim at treating the usual curvature 
form.

The inclusion of a density-varying flow was tackled by Rudman~\cite{Rudman1998}  
in the context of VOF by adopting a consistent mass and momentum transport 
scheme, which was focused on the simulation of multiphase flow with large 
density ratios.
This approach was also adopted by Raessi and Pitch~\cite{Raessi2012} and Ghods 
and Hermann~\cite{Ghods2013} within the level-set method, while the work of 
Mirjalili and Mani~\cite{Mirjalili2021} presented not only a consistent mass and 
momentum transport scheme, but also an energy-preserving scheme in the context 
of phase-field methods.

In this work, the framework of the well-known (mass) Conservative Level-Set 
method \cite{Olsson2005} is adopted for capturing the moving interface.
Base on the energy-preserving level set method introduced in \cite{Valle2020}, 
we produce a fully energy-preserving method for multiphase flows by including a 
consistent mass and momentum transport as in Mirjalili and 
Mani~\cite{Mirjalili2021}.
Equipped with such a physics-compatible discretization, we target DNS of 
vertical falling films.

The rest of the paper is organized as follows: in 
section~\ref{sec_ff:math-model} the governing equations are introduced, in 
section~\ref{sec_ff:num-method} the numerical method is detailed, while in 
section~\ref{sec_ff:results} the cases under consideration are introduced and 
the results commented. Finally, in section~\ref{sec_ff:conclusions} conclusions 
are drawn and future developments sketched.

\section{MATHEMATICAL MODEL}
\label{sec_ff:math-model}

The subtle physical equilibrium at which the fluid system is subject calls for a 
careful, and thus conservative, formulation of the governing equations.

The interface separating the two phases is modeled implicitly by means of a 
marker function $\ls \in [0,1]$ which can be regularized as needed, as will be 
discussed in next section.
The values $\ls = 0$ indicate the liquid phase, $\ls = 1$ corresponds with the 
gaseous one and $\ls = 0.5$ with the interface location.
For a sufficiently well-behaved $\ls$, the interface normal is defined as
\begin{equation}
  \ni = \frac{\nabla\ls}{|\nabla\ls|}
  \label{eqn:multiphase_marker_normal}
\end{equation}
while the interface curvature is defined as
\begin{equation}
  \ki = \div \ni
  \label{eqn:multiphase_marker_curvature}
\end{equation}
Note that when the marker function is the distance function, $|\nabla\ls| = 1$ 
and thus $\ni = \nabla\ls$.

The flow is assumed incompressible for both liquid and gaseous phases
\begin{equation}
  \div \vel= 0
  \label{eqn:Nusselt-divergence}
\end{equation}
under this assumption, the marker function obeys the following transport 
equation
\begin{equation}
  \frac{\partial \ls}{\partial t} + \div \left( \vel \ls \right) = 0
  \label{eqn:multiphase_level_set}
\end{equation}
while the momentum transport is ruled by the conservative version of the 
dimensionless Navier-Stokes equations.
Exploiting the Nusselt flat film solution we introduce Reynolds ($\Reynolds$) 
and Weber ($\Weber$) numbers.
Additionally, because we are concerned with the solution of both liquid and 
gaseous phases simultaneously, we introduce density ($\rd = \rho_g/\rho_l$) and 
viscosity ($\rv = \mu_g/\mu_l$) ratios.
Accordingly, we consider the non-dimensional versions of density ($X_\rho = 
\rho/\rho_l \in [1,\rd]$) and viscosity ($X_\mu = \mu/\mu_l \in [1, \rv]$).
We finally end up with the following momentum equation
\begin{equation}
  3\Reynolds \left( \frac{\partial \left( X_\rho \vel \right)}{\partial t}
                    + \nabla \cdot \left( X_\rho \vel \otimes \vel \right) 
                  \right)
                  = - \nabla p + \nabla \cdot 2 X_\mu \strain + X_\rho\hat{g}
  \label{eqn:Nusselt-momentum}
\end{equation}
subject to capillary forces at the interface, which impose a stress 
discontinuity, $\left[ \stress \right]$ as
\begin{equation}
  \left[ \stress \right]\ni = -\Weber \ki \ni
  \label{eqn:Nusselt-capillarity}
\end{equation}
being $\stress = -p\Id{} + \mu \strain$ the stress tensor, $\strain = 1/2 \left( 
\nabla \vel + (\nabla \vel)^T \right)$ the strain tensor and $\hat{g} = (0, -1, 
0)$.

Which introduce the following dimensionless parameters
\begin{align}
  \Reynolds &= \frac{\rho_l \uN \hN}{\mu_l}
  \label{eqn:Nusselt-Reynolds}\\
  \Weber &= \frac{\st}{\rho_l \uN \hN^2}
  \label{eqn:Nusselt-Weber}
\end{align}
being $\hN$ the undisturbed film thickness and $\uN$ the mean liquid velocity.  
The domain under consideration is assumed to be periodic in both $y$ and $z$ 
directions, while solid walls are present in the $x$ direction.
In a symmetric setup with respect to the $x$ mid-plane, there is a film flowing 
down each wall.

No-slip velocity
\begin{equation}
    \vel\rvert_{wall} = \vec{0}
  \label{eqn:Nussel_no-slip}
\end{equation}
and no marker flow
\begin{equation}
  \nabla \ls \cdot \hat{n}\rvert_{wall} = 0
  \label{eqn:Nussel_no-flow}
\end{equation}
are prescribed at both solid walls.

\section{NUMERICAL METHOD}
\label{sec_ff:num-method}

\subsection{Discretization}
\label{sec_ff:num-method-discretization}
We proceed as in~\cite{Valle2020} by discretizing differential geometry 
operators from a geometrical perspective and then construct discrete vector 
calculus operators within a finite volume method.
Once we obtain the discrete versions of the differential geometry operators, we 
construct the discrete counterparts of divergence~($\DIV$), gradient~($\GRAD$), 
Laplacian~($\LAP$) and convective~($\ADV{\cdot}$) operators, resulting in a 
classical finite volume, second order, staggered method as introduced by Harlow 
and Welch~\cite{Harlow1965, Verstappen2003}.
In addition, we introduce a high resolution advection schemes ($\CONV$) via flux 
limiters in a similar fashion~\cite{Valle2018}.

After adopting a proper regularization,
 we introduce the following equations for the physical properties
\begin{align}
    \sfield[f]{\uprho} &= 1 + \left( \rd - 1 \right)\lsv{f}
    \label{eqn:Xrho-discrete}\\
    \sfield[f]{\upmu} &= 1 + \left( \rv - 1 \right)\lsv{f}
    \label{eqn:Xmu-discrete}
\end{align}
while the original set of governing
equations~(\ref{eqn:Nusselt-divergence}-\ref{eqn:Nusselt-momentum}) is 
discretized as
\begin{align}
  \DIV \uf                      &= \Zv{c}
  \label{eqn:divergence-discrete}\\
  \frac{d\lsv{c}}{dt}           &= - \CONV\lsv{c}
  \label{eqn:marker-discrete}\\
  3\Reynolds\frac{d\left( \RHO\uf \right)}{dt} &=
                                 - \ADV{\RHO\uf}\uf
                                 - \GRAD \pc
                                 + \LAP\uf
                                 + \Weber~\K[F] \GRAD \lsv{c}
                                 + \RHO \vfield{g}
  \label{eqn:momentum-discrete}
\end{align}
where $\uf$ stands for the staggered velocity field, $\lsv{c}$ is the collocated 
marker function and $\pc$ the collocated pressure, $\RHO = 
\mathsf{diag}\left(\sfield[f]{\uprho} \right)$ is the diagonal matrix 
arrangement of the staggered density, and $\K[F] = \mathsf{diag}\left( \E\kc 
\right)$ is the diagonal arrangement of the staggered curvature that was 
introduced in Valle et al.~\cite{Valle2020}.

%

The conservation of mass is a consequence of the conservation of the marker 
function stated in equation~(\ref{eqn:marker-discrete}) and the linear 
reconstruction of density as in equation~(\ref{eqn:Xrho-discrete})
\begin{equation}
    \frac{d\sfield[c]{\uprho}}{dt}
    = (\rho_1-\rho_0)\frac{d\lsv{c}}{dt}
    =-(\rho_1-\rho_0)\CONV\lsv{c}
    =-\CONV\sfield[c]{\uprho}
    \label{eqn:mass-discrete}
\end{equation}
which it is implicitly defined by means of equations~(\ref{eqn:Xrho-discrete}) 
and~(\ref{eqn:marker-discrete}).

\subsection{Conservation of energy}
\label{sec_ff:num-method-discretization-energy}

The evolution of kinetic energy $E_k=1/2\DOT{\rho\vel}{\vel})$ for multiphase 
flows is discussed by first analyzing the conservation of energy in terms of the 
velocity field $\vel$ and the conservative equations~(\ref{eqn:mass-discrete}) 
and~(\ref{eqn:momentum-discrete}) as in~\cite{Mirjalili2021}
\begin{equation}
  \begin{aligned}[b]
    \frac{d E_k}{dt}
      = \frac{1}{2}\DOT{\vel}{\frac{\partial \left( \rho\vel \right)}
                             {\partial t}}
      +  \frac{1}{2}\DOT{\rho\vel}{\frac{\partial \vel}{\partial t}}
      = \DOT{\vel}{\frac{\partial \left( \rho\vel \right)}{\partial t}}
       - \frac{1}{2}\DOT{\vel}{\vel\frac{\partial \rho}{\partial t}}
     \end{aligned}
  \label{eqn:dEkdt}
\end{equation}
where we have included the inner product $\DOT{f}{g} = \int fg dV$, which 
applies to both continuous and discrete fields.
We then obtain the discrete counterpart of equation
equation~(\ref{eqn:dEkdt}) by including equations~(\ref{eqn:momentum-discrete}) 
and~(\ref{eqn:mass-discrete}), the former requiring the use of the isometric 
cell-to-face interpolation operator $\sCF$ in order to match dimensions of the 
discrete field.
After rearranging terms, we obtain
\begin{equation}
  \begin{aligned}[b]
      \frac{d\sfield{E_k}}{dt} =\frac{d\DOT{\uf}{\RHO\uf}}{dt} =
   &- \DOT{\uf}{\ADV{\RHO\uf}\uf}
    + \DOT{\uf}{\GRAD\pc}
    + \DOT{\uf}{\LAP\uf}
    + \Weber\DOT{\uf}{\K[F] \GRAD \lsv{c}}
    + \DOT{\uf}{\RHO \dvec{g}}\\
    &+ \frac{1}{2}\DOT{\uf}{\U\sCF\CONV\sfield[c]{\uprho}}
  \end{aligned}
  \label{eqn:dEkdt-discrete}
\end{equation}
where $\U = diag\left( \uf \right)$ is the diagonal arrangement of the staggered 
velocity field.
Note that the discrete inner product takes the form of a weighted vector dot 
product, i.e., $\DOT{\uf}{\RHO\uf} = \uf^T\mathsf{V_f}\RHO\uf$, where 
$\mathsf{V_f} = diag(v_f)$ is the diagonal arrangement of the staggered volumes.
We have also included the implicit version of the mass transfer equations, even 
when, as it was discussed above, this equation is not explicitly computed.

We now analyze the evolution of potential energy,
\begin{equation}
  \frac{dE_p}{dt} = -\Weber \DOT{\vel}{\ki\grad\ls}
                    -       \DOT{\vel}{X_\rho\vec{g}}
  \label{eqn:dEpdt-regularized}
\end{equation}
which is composed of the capillary and the buoyancy term.
Analogously, the evolution of discrete potential energy equation is obtained in 
an analogous way as in equation~(\ref{eqn:dEpdt-regularized})
\begin{equation}
  \frac{d \sfield{E_p}}{dt}
  = -\Weber \DOT{\uf}{\K[F] \GRAD \lsv{c}} - \DOT{\uf}{\RHO\dvec{g}}
  \label{eqn:dEpdt-discrete}
\end{equation}
where we have adopted an energy-preserving discretization of the capillary terms 
as we did in~\cite{Valle2020}.

Finally, we are now able to assess the evolution of discrete mechanical energy 
by including equations~(\ref{eqn:dEkdt-discrete}) 
and~(\ref{eqn:dEpdt-discrete}).
After rearranging, we obtain
\begin{equation}
    \frac{d\sfield{E_m}}{dt} =
    - \DOT{\uf}{\ADV{\RHO\uf}\uf}
    + \frac{1}{2}\DOT{\uf}{\U\sCF\CONV\sfield[c]{\uprho}}
    + \DOT{\uf}{\LAP\uf} < 0
  \label{eqn:dEmdt-discrete}
\end{equation}
where the pressure terms has vanished given that $\DIV\uf = 0$, as shown 
in~\cite{Verstappen2003}.

We are, however, left with the two first convective terms in 
equation~(\ref{eqn:dEmdt-discrete}) which should, in virtue of its convective 
nature, ideally vanish.
From this point onward, we follow Mirjalili and Mani~\cite{Mirjalili2021} to 
show that the adoption of a consistent mass and momentum transport along with a
smart interpolation strategy results in an energy-consistent discretization.
To do so, we note first that the momentum convective operator is not 
skew-symmetric due to $\DIV\RHO\uf \neq 0$ if density is not constant.
However, due to the mimetic structure adopted for the construction of 
$\ADV{\cdot}$, we can decompose $\ADV{\RHO\uf}$ as
\begin{equation}
    \ADV{\RHO\uf} = \OFFADV{\RHO\uf}
                + \frac{1}{2}diag\left( \sCF \DIV \RHO \uf \right)
  \label{eqn:ADV-decomposition}
\end{equation}
The off-diagonal part $\OFFADV{\RHO\uf}$ is a purely skew-symmetric operator, 
which results into an energy neutral operator, and we are left with the diagonal 
one $1/2diag\left( \sCF\DIV\RHO\uf\right)$, where the $1/2$ factor arise from 
the interpolation applied to the transported velocity 
field~\cite{Verstappen2003}.

Introducing equation~(\ref{eqn:ADV-decomposition}) into 
equation~(\ref{eqn:dEmdt-discrete}) we obtain
\begin{equation}
    \frac{d\sfield{E_m}}{dt} =
    - \frac{1}{2}\DOT{\uf}{diag\left( \sCF \DIV \RHO \uf \right)\uf}
    + \frac{1}{2}\DOT{\uf}{\U\sCF\CONV\sfield{\uprho}}
    +            \DOT{\uf}{\LAP\uf}
  \label{eqn:dEmdt-discrete-decomposition}
\end{equation}
then, exploiting $\mathsf{A}\sfield{b} = diag(\sfield{a})\sfield{b} = 
diag(\sfield{b})\sfield{a} = \mathsf{B}\sfield{a}$ to rearrange 
$diag\left(\sCF\DIV\RHO\uf\right)\uf=\U\sCF\DIV\RHO\uf$ first and $\RHO\uf = 
\U\sfield[f]{\uprho}$ later, we obtain
\begin{equation}
  \begin{aligned}
    \frac{d\sfield{E_m}}{dt} &=
    - \frac{1}{2}\DOT{\uf}{\U\sCF \DIV  \U \sfield[f]{\uprho}}
    + \frac{1}{2}\DOT{\uf}{\U\sCF \CONV    \sfield[c]{\uprho}}
    +            \DOT{\uf}{\LAP\uf}\\
  \end{aligned}
  \label{eqn:dEmdt-discrete-decomposition-1}
\end{equation}
We then introduce the definition $\CONV = \DIV\U\HR$ introduced 
in~\cite{Valle2018,Valle2020}, where $\HR$ contains the high resolution 
cell-to-face interpolation, to yield
\begin{equation}
  \begin{aligned}
    \frac{d\sfield{E_m}}{dt} &=
    - \frac{1}{2}\DOT{\uf}{\U\sCF \DIV  \U     \sfield[f]{\uprho}}
    + \frac{1}{2}\DOT{\uf}{\U\sCF \DIV  \U \HR \sfield[c]{\uprho}}
    +            \DOT{\uf}{\LAP\uf}\\
  \end{aligned}
  \label{eqn:dEmdt-discrete-decomposition-simplified}
\end{equation}
from where we can infer the proper cell-to-face interpolation for the discrete 
density field as
\begin{equation}
  \sfield[f]{\uprho} = \HR \sfield[c]{\uprho}
  \label{eqn:density-interpolation}
\end{equation}
such that convective contributions to the discrete mechanical energy cancel out 
and finally yield
\begin{equation}
    \frac{d\sfield{E_m}}{dt} = \DOT{\uf}{\LAP\uf} < 0
  \label{eqn:dEmdt-discrete-final}
\end{equation}
Note, however, that $\sfield[c]{\uprho}$ is actually not computed but can easily 
be obtained from $\lsv{f}$ as
\begin{equation}
  \sfield[f]{\uprho} = \rho_0 + (\rho_1 - \rho_0)\lsv{f}
  \label{eqn:density-marker-relation}
\end{equation}
stating an explicit relationship between the advection of the marker function 
and the reconstruction of density.
This was previously introduced in the literature, along with other specific 
techniques, as consistent mass and momentum 
transport~\cite{Rudman1998,Raessi2012,Ghods2013,Mirjalili2021}.
In summary, we have developed a consistent discretization that can easily 
preserve mass, momentum (up to surface tension) and energy.
%

Following the original work of Rudman~\cite{Rudman1998}, the adoption of the FSM 
method can proceed with slight modifications.
We adopt the LU decomposition approach introduced by Perot~\cite{Perot1993} in 
order to show the integration procedure of an explicit time integration scheme 
and the modification with respect to the classical FSM followed by 
Rudman~\cite{Rudman1998}.
%
First, following the conservative nature of the governing, equations, we will 
integrate in time discrete momentum i.e., considering $(\RHO\uf)$ as variable on 
its own.
Second, we will use the new density and momentum fields in order to obtain a new 
velocity field which is consistent with the former two.

The semi-discretized equations~(\ref{eqn:marker-discrete}) 
and~(\ref{eqn:momentum-discrete}) can be compiled together with the 
velocity-momentum expression and the usual divergence free constrain $\DIV\uf=0$ 
to result in a linear system of equations as
\begin{equation}
    \begin{pmatrix}
        d_t  &   0      &  0     &   0      \\
        0  &   d_t      &  0     &   \GRAD  \\
        0  &   \Ic{f} &  -\RHO &   0      \\
        0  &   0      &  \DIV  &   0      \\
    \end{pmatrix}
    \begin{pmatrix}
        \lsv{c} \\
        \RHO\uf \\
        \uf     \\
        \pc     \\
    \end{pmatrix}
    =
    \begin{pmatrix}
        -\CONV \lsv{c}                          \\
        \mathsf{R}(\sfield[f]{\uprho}^n, \uf^n, \pc^n)\\
        \Zv{f}                                  \\
        \Zv{c}                                  \\
    \end{pmatrix}
    \label{eqn:gov-matrix}
\end{equation}
where $\mathsf{R}(\sfield[f]{\uprho}^n, \uf^n, \pc^n) = -\ADV{\RHO^n\uf^n}\uf^n 
-\LAP\uf^n + \Weber \K^n \GRAD \lsv{c}^n + \RHO^n\vfield{g}$ corresponds with a 
generalized
explicit treatment of the right hand side of 
equation~(\ref{eqn:momentum-discrete}) without the pressure term.
Note that the left-hand-side is evaluated at time level $n+1$, while the right 
hand side is evaluated at $n$ (or, in general, previous time-steps).
In an explicit setup as the one described here, this implies that the density 
field used for the momentum transport (2nd row) is evaluated at time level $n$, 
whereas the density field used to impose the divergence-free velocity field is 
evaluated at time level $n+1$.
We can then perform an LU decomposition as
\begin{equation}
    \begin{pmatrix}
        d_t  &   0      &  0     &   0      \\
        0  &   d_t      &  0     &   \GRAD  \\
        0  &   \Ic{f} &  -\RHO &   0      \\
        0  &   0      &  \DIV  &   0      \\
    \end{pmatrix}
    =
    \begin{pmatrix}
        d_t       &   0       &   0     & 0   \\
        0       &   d_t       &   0     & 0   \\
        0       &   \Ic{f}  &   -\RHO & 0   \\
        0       &   0       &   \DIV  & -\frac{1}{d_t}\DIV\RHO^{-1}\GRAD \\
    \end{pmatrix}
    \begin{pmatrix}
        \Ic{c}  &   0       &  0      & 0      \\
        0       &   \Ic{f}  & 0       & \frac{1}{d_t}\GRAD \\
        0       &   0       &  \Ic{f} & \frac{1}{d_t}\RHO^{-1}\GRAD      \\
        0       &   0       &  0      & \Ic{c} \\
    \end{pmatrix}
    \label{eqn:LU-decomposition}
\end{equation}
From here, the modified FSM algorithm follows by the usual solution of the 
system by $\mathsf{LU}$ inversion.
We finally obtain the resulting algorithm as
\begin{algorithm}[!h]
  \begin{tabular}{rllll}
    1 & Integrate & $\frac{d\lsv{c}}{dt}$ & $= -\CONV\lsv{c}^n$
                  & $\to \lsv{c}^{n+1}$\\
    2 & Integrate & $\frac{d\left( \RHO\uf \right)}{dt}$ & $= 
                \mathsf{R}(\sfield[f]{\uprho}^n, \uf^n, \pc^n)$
                  & $\to \left( \RHO\uf \right)^{\star}$ \\
    3 & Solve     & $\RHO^{n+1} \uf^\star$ & $ = \left( \RHO\uf \right)$
                  & $\to \uf^\star$\\
    4 & Solve     & $\DIV\RHO^{-1}\GRAD\pc^{n+1}$ & $ = \DIV\uf^\star$
                  & $\to \pc^{n+1}$\\
    5 & Correct   & $\uf^{n+1}$ & $ = \uf^\star - \frac{1}{dt}\RHO^{-1}\GRAD\pc$
                  & $\to \uf^{n+1}$\\
    6 & Update    & $\left( \RHO\uf \right)^{n+1}$ & $ = \RHO^{n+1}\uf^{n+1}$
                  & $\to \left( \RHO\uf \right)^{n+1}$\\
  \end{tabular}
  \caption{Integration of the governing equations along a Fractional Step Method 
  as in~\cite{Rudman1998}}
  \label{alg:FSM}
\end{algorithm}

where we have we have re-stated the computation of the $\left( \RHO\uf 
\right)^{n+1}$ in the equivalent form of step $6$ to reassert that the new 
momentum field is consistent with the divergence-free velocity and density 
fields. We have also redefined pressure to include time integration, as it is
customary. 

\section{PRELIMINARY RESULTS}
\label{sec_ff:results}

With the aforementioned discretization in mind, we showcase the proposed DNS 
discretization of the system with the simulation of several falling films.
Introducing the Kapitza number
\begin{equation}
  \Kapitza = \frac{\st}{\rho_l^{1/3}\mu_l^{4/3}}
  \label{eqn:Kapitza}
\end{equation}
we obtain a dimensionless number that is fixed with the physical properties of 
the working liquid, while the density~($\rd$) and viscosity~($\rv$) ratios 
provide with the information regarding the vapor phase.
In this manner, the liquid-gas pair properties can be fully described by 
$\Kapitza$, $\rd$ and $\rv$, as seen in Table\ref{tab:fluid-properties}, while 
$\Reynolds$ is left as the solely variable controlling the fluid dynamics of the 
system, which i set at $Re = \{150,200\}$ to define typical falling film 
dynamics involved in absorption chillers.
\begin{table}[h]
  \centering
  \begin{tabular}{llrrrrrr}
    id &  fluid         & $T(^{\circ}C)$ &$X_{abs}$  &  $\Kapitza$ & 
    $\rd(\times10^{-3})$   & $\rv(\times10^{-3})$   \\
    \hline
    $A$ & $H_2O$          & $9.9$   &$0.00$ &  $2420$ & $0.60$ & $9.37$ \\
    $B$ & $H_2O/LiBr$     & $50.0$  &$0.60$ &  $443$  & $0.35$ & $2.45$ \\
    $C$ & $H_2O/Carrol$   & $50.0$  &$0.67$ &  $150$  & $0.35$ & $1.10$ \\
  \hline
  \end{tabular}
  \caption{Estimated fluid properties dimensionless groups. For three typical 
  working fluids in an absorption chiller.}
  \label{tab:fluid-properties}
\end{table}

The computational domain is a $[10\hN \times \Lambda_y \times \Lambda_z]$ box 
with periodic boundaries in both $y$ and $z$ directions, where 
$\Lambda_y=\Lambda_x=100\hN$, which correspond with the length of a long-wave 
perturbation.
We will denote $x$, $y$ and $z$ directions as wall-normal, stream-wise and 
span-wise, and $N_x$, $N_y$ and $N_z$ the number of nodes in the $x$, $y$ and 
$z$ directions.
The mesh is refined in the vicinity of the flat film thickness and coarsened at 
the center of the box, where the gas phase is expected.
The following expression gives the refinement introduced in the $x$ axis
\begin{equation}
  \begin{aligned}[b]
  x &= L_x \left(      0.5 f_x \left( 1%
      + \frac{sinh\left(\alpha \left(\frac{i}{N_r} - 0.5\right)\right)}%
             {sinh\left(\frac{\alpha}{2}\right)}
            \right)\right)  &\forall i \in [0, N_r]\\
  x &= L_x\left( f_x + 0.5 (1-2f_x) \left( 1%
      + \frac{tanh\left(\beta \frac{i-N_r}{N_x-2N_r} - 0.5  \right)}%
             {tanh\left(\frac{\beta}{2}\right)}
            \right)\right)  &\forall i \in [N_r, N_x - N_r]\\
  x &= L_x\left( (1-f_x) + 0.5 f_x \left( 1%
      + \frac{sinh\left(\alpha \frac{i-N_x+N_r}{N_r} - 0.5  \right)}%
             {sinh\left(\frac{\alpha}{2}\right)}
         \right)\right)     &\forall i \in [N_x-N_r, N_x]\\
  \end{aligned}
  \label{eqn:x-position}
\end{equation}
where $L_x = 10$, $f_x=0.2$ is the fraction of $L_x$ refined close to the wall, 
while $N_r = N_x/3$ is the number of nodes introduced in the refinement regions.
Parameters $\alpha = 2$ and $\beta=2$ control the smoothness of the refinements.

We set up a symmetric layout consisting of two films falling down parallel 
walls.
The interface is initialized as in Dietze et al.~\cite{Dietze2014} by 
prescribing a perturbation at the film surface as
\begin{equation}
  h = \hN\left( 1
    + \epsilon_y cos\left(2\pi\frac{y}{\Lambda_y}\right)
    + \epsilon_z cos\left(2\pi\frac{z}{\Lambda_z}\right) \right)
    \label{eqn:film-shape}
\end{equation}
where $h$ is the perturbed film thickness, $\epsilon_y = 0.2$ and $\epsilon_z = 
0.05$ are chosen according to Dietze et al.~\cite{Dietze2014} and correspond 
with amplitude of the perturbation in the $y$ and $z$ directions, respectively.
The wavelength $\Lambda_y$ and $\Lambda_z$ are sufficiently large to represent 
long-wave perturbations~\cite{Kalliadasis2012} and
the velocity profile is initialized to the undisturbed flat falling film 
solution.

Time integration is performed with a 3rd order Runge-Kutta method for step $1$ 
and a 2nd order Adams-Bashforth one for step $2$.
The solver used for the Poisson equation is a Preconditioned Conjugate Gradient 
(PCG) method preconditioned with $\RHO$ in order to reduce the condition number 
of the system.

Results shown in figures~1 and~2 show the impact of decreasing Kapitza number 
into the dynamics of the surface.
The stabilizing effect of surface tension fades as the Kapitza number is 
reduced, thus enhancing the appearance of larger humps and complex 
instabilities.

In this regard, results show the appearance of a leading depression before the 
arrival of the wave tip, which drains fluid in the $z$ direction, which 
contributes to the appearance of $3D$ structures as shown in~\cite{Dietze2014}.
This depression is more pronounced in the low Kapitza cases due to the 
stabilizing effect that capillary force plays in the development of surface 
deformations.
While the stiffer case $A$ presents a less marked curvature, it also presents 
more undulations on the film surface.
This results in the appearance of a more dispersed velocity field on top of the 
film surface.
Conversely, lower Kapitza numbers result in more acute film deformations, which, 
on the other hand, contribute to smaller, but also more coherent, flow patterns.

\foreach \Re in {150,200}
{
  \begin{figure}[!htb]
    \centering
    \foreach \f in {A,B,C}
    {
      \begin{minipage}{0.3\textwidth}
        \centering
        \includegraphics[width=\textwidth]{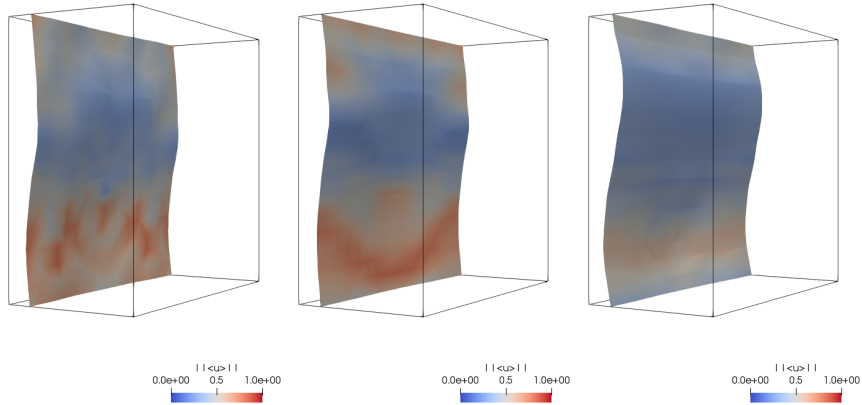}
      \end{minipage}
    }
    \label{fig:stresses-\Re}
    \caption{Velocity magnitude on the film surfaces for cases $A$, $B$ and $C$ 
  at $\Reynolds = \Re$ after $T = 100$.}
  \end{figure}
}

Regarding the shape of the film interface, it can be observed how case $B$ 
produces a more abrupt hump, showing a marked preceding wave rise and a also a 
central delay on wave maximum, pushing the flow in the $z$ direction and thus 
revealing the incipient formation of a horseshoe pattern.
On the other hand, milder surface effects in case $C$ result in a smoother 
undulation, which show smaller $z$ axis flow which is mainly in the $y$ 
direction.
These effects are intensified with the increase of the Reynolds number, as it 
can be seen for case $B$ at $\Reynolds = 200$.
In that situation, the wave is rolling on top of itself thus increasing the hump 
height an resulting into a growing large scale instability.
On the other hand, the high Kapitza case presented in case $A$ shows milder 
velocity fields at higher Reynolds numbers, which may caused by an intensified 
momentum diffusion at the interface due to the high frequency pattern that the 
undulations on top of the surface form.

%

\section{CONCLUSIONS}
\label{sec_ff:conclusions}

The method has been deployed in the DNS of vertical falling films under extreme
density ratios, which pose a major numerical challenge and at the same time are
of industrial relevance.
Results are in accordance with the expected behavior of such films and reported
in both experimental and numerical literature.  In this regard, it is shown
capillary forces stabilizing effect on the film dynamics.  Accordingly,
reducing surface tension (e.g., by adding surfactants) is observed to enhance
instability of the system.

Extending our previous work on the energy-preserving inclusion of
curvature~\cite{Valle2020} and adopting the ideas presented in Mirjalili and
Mani~\cite{Mirjalili2021} for the discretization of the convective terms, we
present a formulation which is mathematically consistent.
The aforementioned merits benefit from the adoption of our heavily
algebra-based formulation both for the discretization and the formulation of
the modified FSM as proposed by Rudman~\cite{Rudman1998}, revealing the close
connection between consistent mass and momentum transfer and the conservation
of energy.

%
While we succeeded at the formulation of a fully mass conservative and
energy-preserving scheme, the conservation of linear momentum is still
unclear~\cite{Kim2005}.  However, as it was already commented
in~\cite{Valle2020}, while the lack of total momentum conservation is an
undesired property, the adoption of an energy-preserving scheme provides a
bound on total energy and thus stability. Nonetheless, the conservation of linear
momentum is an active line of research that deserves further discussion.

\section*{ACKNOWLEDGEMENTS}
\label{sec_ff:acknowledgments}

The authors acknowledge the computing resources provided by the Barcelona 
Supercomputing Center under contracts IM-2019-3-0026 and IM-2020-1-0006. An also grants from the \textit{Ministerio de Economía y Competitividad, Spain} under contracts
ENE2015-70672-P and ENE2017-88697-R.
N. Valle also acknowledges an FI AGAUR-Generalitat de Catalunya fellowship under 
contract 2017FI\_B\_00616.

\bibliographystyle{unsrt}
\bibliography{/home/nvm/res/ref/library.bib}

\end{document}